\documentclass[a4paper,11pt]{article}
\usepackage{pos}
\usepackage{float}
\usepackage{caption}
\usepackage{subcaption}

\usepackage[backend=bibtex, natbib=true,style=numeric,sorting=none,backref=true]{biblatex}
\usepackage{lineno}

\title{Decelerated sub-relativistic material with energy Injection}
 \ShortTitle{Decelerated sub-relativistic material with energy Injection}

\author*[a,b]{B. Betancourt Kamenetskaia}
\author[c]{N. Fraija}
\author[d,e,f,g]{M. Dainotti}
\author[c]{A. Gálvan-Gámez}
\author[h]{R. Barniol Duran}
\author[i,j]{S. Dichiara}

\affiliation[a]{TUM Physics Department, Technical University of Munich, James-Franck-Straße, 85748 Garching, Germany}
\affiliation[b]{LMU Physics Department, Ludwig Maxmillians University, Theresienstr. 37, 80333 Munich, Germany}
\affiliation[c]{Instituto de Astronom\'ia, Universidad Nacional Aut\'onoma de M\'exico, Ciudad de México, México}
\affiliation[d]{Physics Department, Stanford University, 382 Via Pueblo Mall, Stanford, USA}
\affiliation[e]{Space Science Institute, Boulder, CO, USA}
\affiliation[f]{Obserwatorium Astronomiczne, Uniwersytet Jagielloński, ul. Orla 171, 31-501 Kraków, Poland}
\affiliation[g]{Interdisciplinary Theoretical \& Mathematical Science Program, RIKEN(iTHEMS), 2-1 Hirosawa, Wako, Saitama, 351-0198, Japan}
\affiliation[h]{Department of Physics and Astronomy, California State University, Sacramento, 6000 J Street, Sacramento, CA 95819-6041, USA}
\affiliation[i]{Department of Astronomy, University of Maryland, College Park, MD 20742-4111, USA}
\affiliation[j]{Astrophysics Science Division, NASA Goddard Space Flight Center, 8800 Greenbelt Road, Greenbelt, MD 20771, USA}

\emailAdd{boris\_betancourt@ciencias.unam.mx}
\emailAdd{nifraija@astro.unam.mx}

\abstract{We investigate the evolution of the afterglow produced by the deceleration of the non-relativistic material due to its surroundings. The ejecta mass is launched into the circumstellar medium with equivalent kinetic energy expressed as a power-law velocity distribution $E\propto (\Gamma\beta)^{-\alpha}$. The density profile of this medium follows a power law $n(r)\propto r^{-k}$ with $k$ the stratification parameter, which accounts for the usual cases of a constant medium  ($k=0$) and a wind-like medium ($k=2$). A long-lasting central engine, which injects energy into the ejected material as ($E\propto t^{1-q}$) was also assumed. With our model, we show the predicted light curves associated with this emission for different sets of initial conditions and notice the effect of the variation of these parameters on the frequencies, timescales and intensities. The results are discussed in the Kilonova scenario.}

\FullConference{37$^{\rm{th}}$ International Cosmic Ray Conference (ICRC 2021)\\
		July 12th -- 23rd, 2021\\
		Online -- Berlin, Germany}

\begin{document}

\def\aj{AJ}
\def\actaa{Acta Astron.}
\def\araa{ARA\&A}
\def\apj{The Astrophysical Journal}
\def\apjl{ApJ}
\def\apjs{ApJS}
\def\ao{Appl.~Opt.}
\def\apss{Ap\&SS}
\def\aap{A\&A}
\def\aapr{A\&A~Rev.}
\def\aaps{A\&AS}
\def\azh{AZh}
\def\baas{BAAS}
\def\bac{Bull. astr. Inst. Czechosl.}
\def\caa{Chinese Astron. Astrophys.}
\def\cjaa{Chinese J. Astron. Astrophys.}
\def\icarus{Icarus}
\def\jcap{Journal of Cosmology and Astroparticle Physics}
\def\jrasc{JRASC}
\def\mnras{MNRAS}
\def\memras{MmRAS}
\def\na{New A}
\def\nar{New A Rev.}
\def\pasa{PASA}
\def\pra{Phys.~Rev.~A}
\def\prb{Phys.~Rev.~B}
\def\prc{Phys.~Rev.~C}
\def\prd{Phys.~Rev.~D}
\def\pre{Phys.~Rev.~E}
\def\prl{Physical Review Letters}
\def\pasp{PASP}
\def\pasj{PASJ}
\def\qjras{QJRAS}
\def\rmxaa{Rev. Mexicana Astron. Astrofis.}
\def\skytel{S\&T}
\def\solphys{Sol.~Phys.}
\def\sovast{Soviet~Ast.}
\def\ssr{Space~Sci.~Rev.}
\def\zap{ZAp}
\def\nat{Nature}
\def\iaucirc{IAU~Circ.}
\def\aplett{Astrophys.~Lett.}
\def\apspr{Astrophys.~Space~Phys.~Res.}
\def\bain{Bull.~Astron.~Inst.~Netherlands}
\def\fcp{Fund.~Cosmic~Phys.}
\def\gca{Geochim.~Cosmochim.~Acta}
\def\grl{Geophys.~Res.~Lett.}
\def\jcp{J.~Chem.~Phys.}
\def\jgr{J.~Geophys.~Res.}
\def\jqsrt{J.~Quant.~Spec.~Radiat.~Transf.}
\def\memsai{Mem.~Soc.~Astron.~Italiana}
\def\nphysa{Nucl.~Phys.~A}
\def\physrep{Phys.~Rep.}
\def\physscr{Phys.~Scr}
\def\planss{Planet.~Space~Sci.}
\def\procspie{Proc.~SPIE}
\let\astap=\aap
\let\apjlett=\apjl
\let\apjsupp=\apjs
\let\applopt=\ao

\maketitle
\section{Introduction}

Long-duration gamma-ray bursts \citep[lGRBs; $T_{90}\gtrsim 2\,{\rm s}$;][]{1993ApJ...413L.101K} are linked to supernovae \citep[SNe;][]{1999Natur.401..453B, 2006ARA&A..44..507W} caused by the core collapse (CC) of dying massive stars \citep{1993ApJ...405..273W, 1998Natur.395..670G}. Short-duration gamma-ray bursts, on the other hand, are linked to the coalescence of binary compact objects (NS-NS or BH-NS)\footnote{NS corresponds to neutron star and BH to black hole.} that result in kilonovae\footnote{A fairly isotropic thermal transient powered by the radioactive decay of rapid neutron capture process nuclei and isotopes} (KNe) \citep[sGRBs; $T_{90}\lesssim 2\,{\rm s}$;][]{1998ApJ...507L..59L, 2005ApJ...634.1202R, 2010MNRAS.406.2650M, 2013ApJ...774...25K, 2017LRR....20....3M}.

It is thought that enormous volumes of materials with a wide range of velocities are propelled in both cases. Non-relativistic ejecta masses such as dynamical ejecta, cocoon material, shock breakout material, and wind ejecta are propelled in NS-NS mergers \citep[e.g., see][]{2011ApJ...738L..32G, 2013ApJ...778L..16H,2013ApJ...773...78B, 2014ApJ...789L..39W} with velocities in the range $0.03\lesssim \beta\lesssim 0.8$ (expressed, hereafter,  in units of the speed of light). Similarly, several ejecta masses with non-relativistic velocities smaller than $\beta\lesssim0.3$ have also been observed in the context of CC-SNe. 

The interaction of the ejecta mass with the surrounding circumburst medium was suggested to characterize multi-wavelength afterglow observations on time scales ranging from days to many years in the non-relativistic domain \citep[e.g., see][]{1997MNRAS.288L..51W, 1999ApJ...519L.155D, 1999MNRAS.309..513H, 2000ApJ...538..187L, 2003MNRAS.341..263H, 2013ApJ...778..107S, 2015MNRAS.454.1711B}. Several authors, \citep[e.g see ][]{2014MNRAS.437.1821M, 2016ApJ...831..141F, 2020ApJ...890..102L, 2020arXiv200607434S, 2019ApJ...871..123F}, took into account the material launched during the coalescence of binary compact objects and computed the synchrotron emission in the radio bands. The authors assumed the existence of a free-coasting phase before the Sedov-Taylor expansion. Tan et al.  \cite{2001ApJ...551..946T} hypothesized that the shock wave's kinetic energy may be characterized by a power-law (PL) velocity distribution. Ever since, several authors  have proposed that the material ejected during binary compact object coalescence and the CC-SNe be characterized by a PL velocity distribution \citep[e.g., see][]{2013ApJ...773...78B, 2014MNRAS.437L...6K, 2015MNRAS.450.1430H, 2015MNRAS.448..417B, 2017LRR....20....3M, 2018Natur.554..207M, 2019ApJ...871..200F, 2019MNRAS.487.3914K, 2019LRR....23....1M, 2012ApJ...750...68L, 2013ApJ...778...63H, 2013ApJ...778...18M, 2014ApJ...797..107M,2020ApJ...896...25F}. 

In this proceedings, we provide a theoretical model that predicts the late-time multi-wavelength afterglow emission created by the deceleration of the outermost non-relativistic ejecta mass in a circumstellar medium. We assume interaction with a medium parametrized by a power law density profile $n(r)\propto r^{-k}$. We also express the equivalent kinetic energy of the outermost matter as a power-law velocity distribution $E\propto \left(\Gamma\beta\right)^{-\alpha}$. Finally, we consider a long-lasting central engine with the kinetic energy as a power-law distribution $E\propto t^{1-q}$.

The ejecta mass begins to decelerate after a long time, when the swept up quantity of material is similar to the ejected mass.
Electrons are accelerated in forward shocks and cooled down by synchrotron radiation during this stage.  We present the predicted synchrotron light curves for ${\rm k}=0$, $1$, $1.5$, $2$ and $2.5$ that cover several ejecta masses launched during the coalescence of binary compact objects and the CC-SNe.

The paper is organized as follows: In Section 2, we introduce the theoretical model that predicts the multi-wavelength afterglow emission generated by the deceleration of the non-relativistic ejecta mass.  

We assume  for the cosmological constants a spatially flat universe $\Lambda$CDM model with  $H_0=69.6\,{\rm km\,s^{-1}\,Mpc^{-1}}$, $\Omega_{\rm M}=0.286$ and  $\Omega_\Lambda=0.714$ \citep{2016A&A...594A..13P}. Prime and unprimed quantities are used for the comoving and observer frames, respectively.

\section{Theoretical Model}\label{}	
We model the afterglow as two components. The first one is the nonrelativistic ejecta mass ($\Gamma\approx1$). The second one is as the first one with the addition of the energy injection model:

\begin{equation}
\label{energy}
E_{2}(t)=\tilde{E}_{\textrm{inj}} t^{1-q}.
\end{equation}

The first component is assumed to follow a spherical profile of the form \cite{2021ApJ...907...78F}

\begin{equation}
\label{profile}
E_{1}=\tilde{E}\left(\Gamma\beta\right)^{-\alpha}\approx\tilde{E}\beta^{-\alpha},
\end{equation}

where $\beta$ is the shock front's velocity and $\Gamma\sim1$ its Lorentz factor. In the nonrelativistic regime, the ejecta mass is described by the Sedov–Taylor solution. Then, the velocity can be written as

\begin{equation}
    \beta=\beta^0 \,\left(1+z\right)^{\frac{3-k}{5-k}}\,A^{-\frac{1}{5-k}}_{\rm k}\,E_{T}^{\frac{1}{5-k}}\, t^{\frac{k-3}{5-k}}\,,
\end{equation}

where $E_{T}=E_{1}+E_{2}$ with $E_1$ and $E_2$ are given in Eqs. (\ref{energy}) and (\ref{profile}) and where the density parameter is denoted by $A_{k}$ and the redshift as $z$. Therefore, the Sedov–Taylor solution can be written in a general form as
%
%

\begin{equation}
    \tilde{E}\beta^{-\alpha}+\tilde{E}_{\textrm{inj}} t^{1-q}={\beta^0}^{k-5}\left(1+z\right)^{k-3}\,A_{\rm k}\,\beta^{5-k}\,t^{3-k}\,.
\end{equation}

With this in mind, we have two limiting cases:

\begin{equation}
    \begin{cases}
    \tilde{E}\beta^{-\alpha}\propto\left(1+z\right)^{k-3}\,A_{\rm k}\,\beta^{5-k}\,t^{3-k},\hspace{0.9cm} \tilde{E}>>\tilde{E}_{\textrm{inj}}, \cr
    \tilde{E}_{\textrm{inj}} t^{1-q}\propto\left(1+z\right)^{k-3}\,A_{\rm k}\,\beta^{5-k}\,t^{3-k},\hspace{0.6cm}  \tilde{E}<<\tilde{E}_{\textrm{inj}}.
    \end{cases}
\end{equation}

Each limiting case leads to a different velocity; they are given by:

\begin{equation}
    \begin{cases}
    \beta\propto\left(1+z\right)^{-\frac{k-3}{\alpha+5-k}}\,A_{\rm k}^{-\frac{1}{\alpha+5-k}}\,\tilde{E}^{\frac{1}{\alpha+5-k}}\,t^{\frac{k-3}{\alpha+5-k}},\hspace{0.5cm} \tilde{E}>>\tilde{E}_{\textrm{inj}}, \cr
    \beta\propto\left(1+z\right)^{\frac{k-3}{k-5}}\,A_{\rm k}^{\frac{1}{k-5}}\,\tilde{E}_{\textrm{inj}}^{-\frac{1}{k-5}}\,t^{\frac{q+2-k}{k-5}},\hspace{1.6cm}  \tilde{E}<<\tilde{E}_{\textrm{inj}}.
    \end{cases}
\end{equation}

Both cases may be written with just one expression:

\begin{equation}\label{beta_dec}
    \beta=\beta^0 \,\left(1+z\right)^{-\frac{k-3}{\alpha+5-k}}\,A^{-\frac{1}{\alpha+5-k}}_{\rm k}\,\tilde{E}^{\frac{1}{\alpha+5-k}}\, t^{\frac{k-(q+2)}{\alpha+5-k}}\,,
\end{equation}

where the case $\tilde{E}>>\tilde{E}_{\textrm{inj}}$ is obtained by setting $q=1$, while the case $\tilde{E}<<\tilde{E}_{\textrm{inj}}$ is obtained by setting $\alpha=0$ and $\tilde{E}=\tilde{E}_{\textrm{inj}}$. With this, we may also write the blast wave radius $r\propto\left(1+z\right)^{-1}\beta t$ as:

\begin{equation}\label{R_dec}
    r=r^0\,\left(1+z\right)^{-\frac{\alpha+2}{\alpha+5-k}}\,A^{-\frac{1}{\alpha+5-k}}_{\rm k}\, \tilde{E}^{\frac{1}{\alpha+5-k}}\,t^{\frac{\alpha+3-q}{\alpha+5-k}}\,.
\end{equation}

\paragraph{Synchrotron emission}
We assume an electron distribution described as $dN/d\gamma_{e}\propto\gamma_{e}^{-p}$ for $\gamma_{e}\geq\gamma_{m}$, where $p$ is the index of the elctron distribution and $\gamma_{m}$ is the Lorentz factor of the lowest-energy electrons. During the deceleration phase, the electron Lorentz factors and the post-shock magnetic field evolve as, $\gamma_{\rm m}\propto t^{\frac{2(k-(q+2))}{\alpha+5-k}}$,  $\gamma_{\rm c}\propto t^{\frac{-1+2q-k(-2+q-\alpha)-\alpha}{\alpha+5-k}}$, and
$B'\propto \,t^{-\frac{-4-k-2q+kq-k\alpha}{2(\alpha+5-k)}}$, respectively.  The corresponding synchrotron break frequencies and max flux density are given by:

\begin{equation}\label{nu_syn_de}
    \begin{aligned}
    \nu^{\rm syn}_{\rm m}&\propto\,\left(1+z\right)^{\frac{20+k(\alpha-6)-2\alpha }{2(\alpha+5-k)}}\,\epsilon^2_{\rm e}\,\epsilon^\frac12_{\rm B}\,  A^{\frac{\alpha-5}{2(\alpha+5-k)}}_{\rm k}\,\tilde{E}^{\frac{10-k}{2(\alpha+5-k)}}\,t^{-\frac{10(2+q)+k(-7-q+\alpha)}{2(\alpha+5-k)}}\cr
    \nu^{\rm syn}_{\rm c}&\propto\,\left(1+z\right)^{-\frac{8-2\alpha + k(3\alpha+2)}{2(\alpha +5-k)}}\, \epsilon^{-\frac32}_{\rm B}\, (1+Y)^{-2} \,A^{-\frac{3(\alpha+3)}{2(\alpha+5-k)}}_{\rm k}\tilde{E}^{\frac{3(k-2)}{2(\alpha+5-k)}}\,t^{-\frac{8-6q+k(-7+3q-3\alpha)+4\alpha}{2(\alpha+5-k)}}\cr
    F^{\rm syn}_{\rm \nu,max}&\propto\,\left(1+z\right)^{\frac{4+2k-4\alpha+3k\alpha}{2(\alpha+5-k)}}\, \epsilon^{\frac12}_{\rm B}\, D_{\rm z}^{-2}\, A^{\frac{3\alpha+7}{2(\alpha+5-k)}}_{\rm k}\,\tilde{E}^{\frac{8-3k}{2(\alpha+5-k)}}\,t^{\frac{14-8q+k(-7+3q-3\alpha)+6\alpha}{2(\alpha+5-k)}},
    \end{aligned}
\end{equation}

where $Y$ is the Compton parameter, $D_{z}$ is the luminosity distance, $\epsilon_{e}$ is the fraction of the shock's thermal energy density that is transmitted to the electrons and $\epsilon_{B}$ is the fraction turned into magnetic energy density \citep{2013ApJ...778..107S}.

Using the synchrotron break frequencies and the spectral peak flux density (eq.~\ref{nu_syn_de}),  the synchrotron light curves in the fast- and slow-cooling regime are:

\begin{equation}\label{fc_dec}
    F^{\rm syn}_{\nu}\propto
    \begin{cases}
    t^{\frac{25-15q+2k(-7+3q-3\alpha)+11 \alpha}{3(\alpha+5-k)}}\, \nu^{\frac13},\hspace{3.1cm}   \nu<\nu^{\rm syn}_{\rm c}, \cr
    t^{\frac{20-10q+k(-7+3q-3\alpha)+8\alpha}{4(\alpha+5-k)}}\, \nu^{-\frac{1}{2}},\hspace{3.1cm} \nu^{\rm syn}_{\rm c}<\nu<\nu^{\rm syn}_{\rm m},\,\,\,\,\, \cr
    t^{-\frac{10p(2+q)-8(5+\alpha)+kp(-7-q+\alpha)+2k(7-q+\alpha)}{4(\alpha+5-k)}}\,\nu^{-\frac{p}{2}},\,\,\hspace{0.8cm}   \nu^{\rm syn}_{\rm m}<\nu\,, \cr
    \end{cases}
\end{equation}

and

\begin{equation}\label{sc_dec1}
    F^{\rm syn}_{\nu}\propto
    \begin{cases}
    t^{\frac{31-7q+2k(-7+2q-2\alpha)+9\alpha}{3(\alpha+5-k)}}\, \nu^{\frac13},\hspace{3.5cm}  \nu<\nu^{\rm syn}_{\rm m}, \cr
    t^{-\frac{10p(2+q)+6(-8+q-2\alpha)+k(21-5q+5\alpha+p(-7-q+\alpha))}{4(\alpha+5-k)}}\, \nu^{-\frac{p-1}{2}},\hspace{0.3cm} \nu^{\rm syn}_{\rm m}<\nu<\nu^{\rm syn}_{\rm c},\,\,\,\,\, \cr
    t^{-\frac{10p(2+q)-8(5+\alpha)+kp(-7-q+\alpha)+2k(7-q+\alpha)}{4(\alpha+5-k)}}\,\nu^{-\frac{p}{2}},\,\,\,\,\hspace{0.9cm}   \nu^{\rm syn}_{\rm c}<\nu\,, \cr
    \end{cases}
\end{equation}

respectively.

\section{Results and discussion}

\begin{figure}[H]
  \begin{subfigure}[b]{0.3\textwidth}
    \includegraphics[width=\textwidth]{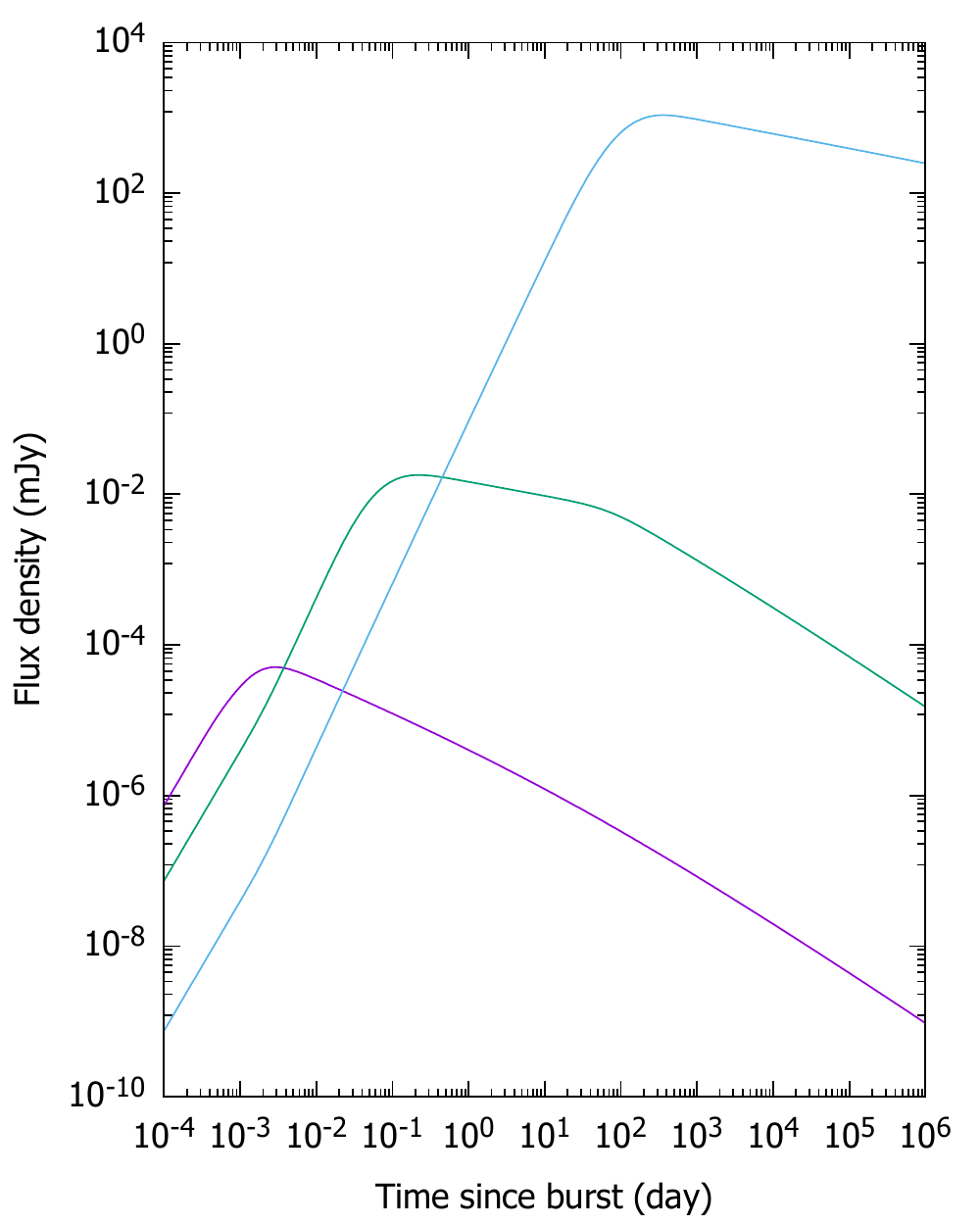}
    \caption{ISM $k=0$.}
    \label{fig:f1}
  \end{subfigure}
  \begin{subfigure}[b]{0.3\textwidth}
    \includegraphics[width=\textwidth]{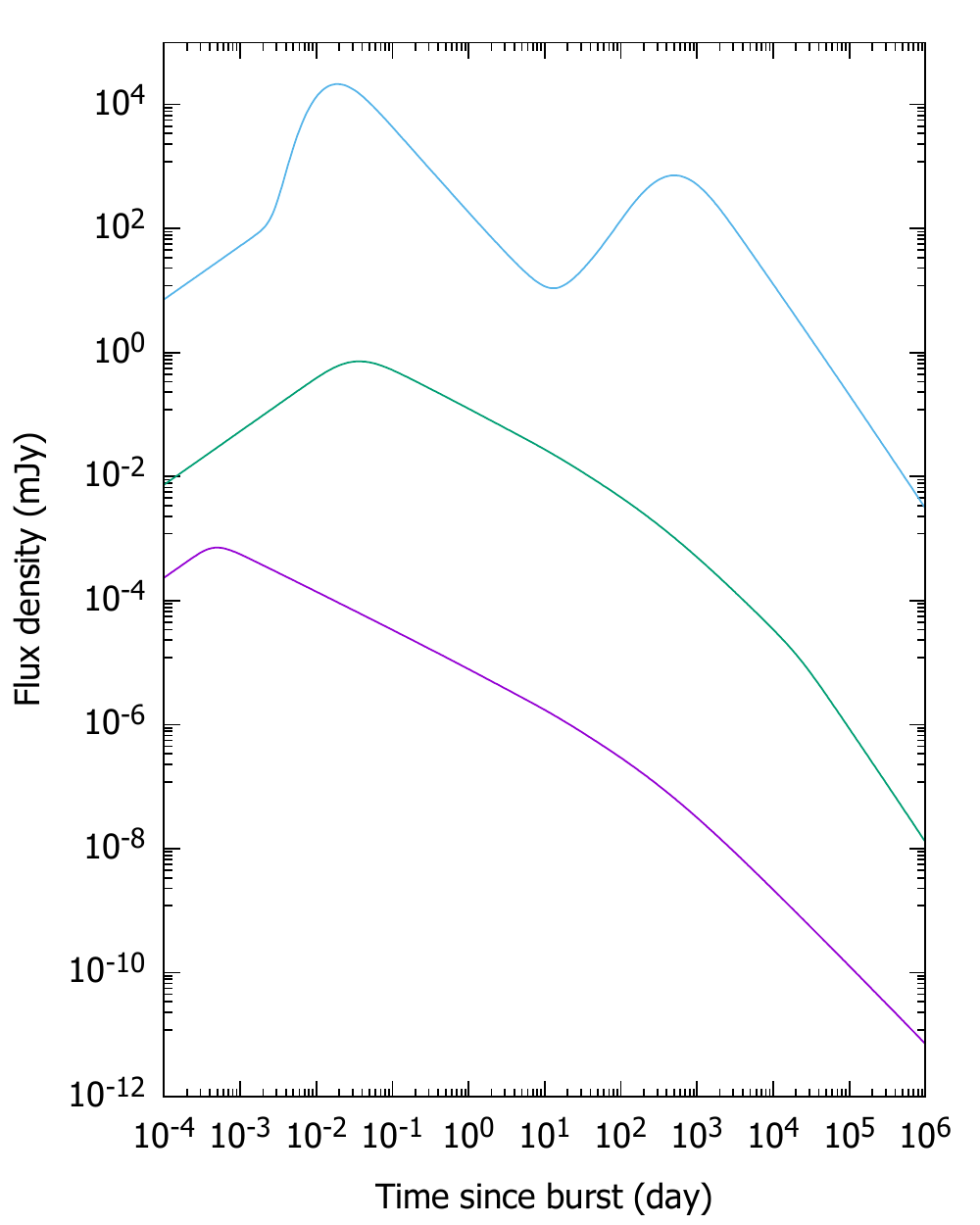}
    \caption{Wind $k=2$}
    \label{fig:f2}
  \end{subfigure}
  \begin{subfigure}[b]{0.3\textwidth}
    \includegraphics[width=\textwidth]{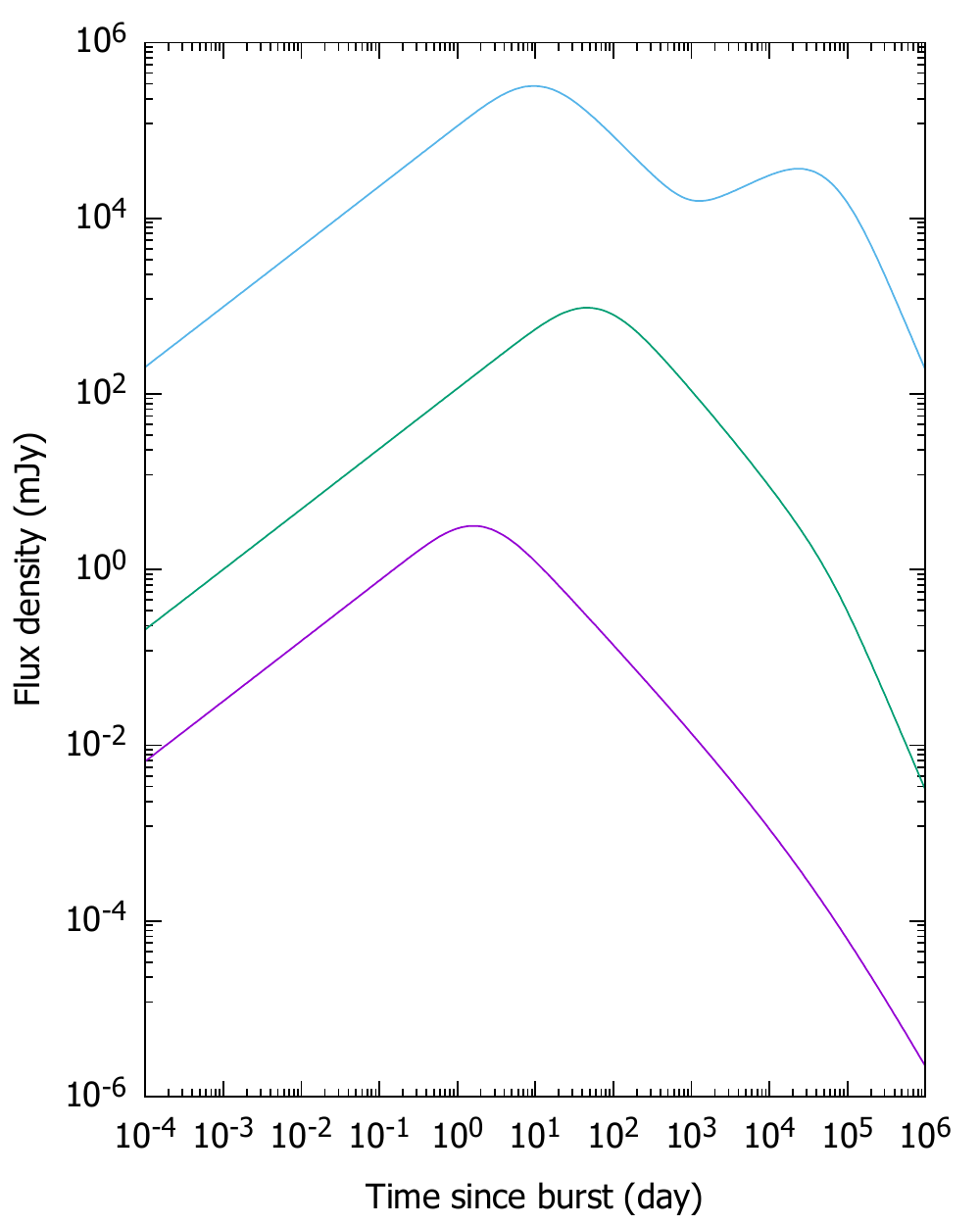}
    \caption{Wind $k=2$ and $q=0.8$.}
    \label{fig:f3}
  \end{subfigure}
  \caption{Synchrotron light curves from a sub-relativistic material decelerated in different circumburst media.}
  \label{LightCurves}
\end{figure}

Figure \ref{LightCurves} shows examples of synchrotron light curves with this model. The purple curves stand for X-ray (1 keV), the green ones for optical (1 eV) and the blue ones for radio (1.6 GHz). Panels \ref{fig:f1} and \ref{fig:f2} correspond to a model without energy injection in an ISM and wind-like medium, respectively. Panel \ref{fig:f3} shows the curves in a wind-like medium for the energy injection component. In all cases, an increase of the stratification parameter leads to a flattening of the rising flux and a steepening of the decreasing one. In the case of radio band, it also leads to the appearance of a double peak. By comparing the first two panels with the third one, it can be seen that the energy injection model leads to larger fluxes by about two orders of magnitude. The right panel also shows that the decrease of the flux starts at much later times than in models without energy injection.

The parameters used for Figure \ref{fig:f1} are: $\tilde{E}=10^{51}\,{\rm erg}$, $\epsilon_{\rm B}=10^{-2}$, $\epsilon_{\rm e}=10^{-1}$, $A_{k}=1\ \mathrm{cm}^{-3}$, $p=2.8$, $q=1.0$, $\alpha=3.0$, $D_z=100\,{\rm Mpc}$ and $z=0.022$. 

For Figure \ref{fig:f2}, they are: $\tilde{E}=10^{51}\,{\rm erg}$, $\epsilon_{\rm B}=10^{-2}$, $\epsilon_{\rm e}=10^{-1}$, $A_{k}=3\times10^{36}\ \mathrm{cm}^{-1}$, $p=2.8$, $q=1.0$, $\alpha=3.0$, $D_z=100\,{\rm Mpc}$ and $z=0.022$.

Finally, for Figure \ref{fig:f3}, they are: $\tilde{E}=10^{51}\,{\rm erg}$, $\epsilon_{\rm B}=10^{-2}$, $\epsilon_{\rm e}=10^{-1}$, $A_{k}=3\times10^{36}\ \mathrm{cm}^{-1}$, $p=2.6$, $q=0.8$, $\alpha=0$, $D_z=100\,{\rm Mpc}$ and $z=0.022$.

\section{Conclusion}
We have derived a model that describes the non-relativistic, adiabatic evolution of the forward shock described by the Sedov-Taylor solution. We have modeled the afterglow in two components: one that considers a long-lasting central engine that injects energy into the shock, which leads to the power-law dependence $E\propto\tilde{E}t^{1-q}$; and one that doesn't assume energy injection and instead assumes a power-law energy distribution $E\propto(\Gamma\beta)^{-\alpha}$. We have also taken into account that the ejecta interacts with a medium parametrized by a power law number density distribution $A\propto R^{-k}$. This general approach is advantageous, as it allows one to not only consider a homogeneous medium ($k=0$) and a wind-like medium ($k=2$), but regions with non-standard stratification parameters, in particular $k=1$, $1.5$ or $2.5$. It also allows one to transition between energy injection models to models without it by the change of two parameters: $\alpha$ and $\beta$.

We have calculated, for both components, the synchrotron light curves in the fast- and slow-cooling regimes and we have analyzed their behaviour for different sets of parameters. In the case of variation of the stratification parameter, we have noticed that an increase of this parameter leads to flatter profiles when the flux increases. It also leads to the appearance of a double-peak behavior in the radio band. As for the case of the comparison between energy injection with models without it, we have shown that the flux increases in general and the moment when it begins to decrease happens later when energy injection is considered.

\acknowledgments{
We acknowledge the support from Consejo Nacional de Ciencia y Tecnolog\'ia (CONACyT), M\'exico, grants IN106521.
}		

\printbibliography


\end{document}